\documentclass[sigconf]{acmart}

\usepackage{subcaption}
\usepackage{float}
\usepackage{graphicx}
\usepackage{wrapfig}
\usepackage{enumitem}
\makeatletter
\newcommand{\setword}[2]{#1\def\@currentlabel{\unexpanded{#1}}\label{#2}}
\makeatother

\copyrightyear{2022}
\acmYear{2022}
\setcopyright{acmlicensed}
\acmConference[MuC '22]{Mensch und Computer 2022}{September 4--7, 2022}{Darmstadt, Germany}
\acmBooktitle{Mensch und Computer 2022 (MuC '22), September 4--7, 2022, Darmstadt, Germany}
\acmPrice{15.00}
\acmDOI{10.1145/3543758.3543761}
\acmISBN{978-1-4503-9690-5/22/09}

\begin{document}

\title[The Human in the \emph{Infinite Loop}]{The Human in the \emph{Infinite Loop}: A Case Study on Revealing and Explaining Human-AI Interaction Loop Failures}

\author{Changkun Ou}
\orcid{0000-0002-4595-7485}
\affiliation{
  \institution{LMU Munich}
  \country{Germany}
}
\email{research@changkun.de}

\author{Daniel Buschek}
\orcid{0000-0002-0013-715X}
\affiliation{
  \institution{University of Bayreuth}
  \country{Germany}
}
\email{daniel.buschek@uni-bayreuth.de}

\author{Sven Mayer}
\orcid{0000-0001-5462-8782}
\affiliation{
  \institution{LMU Munich}
  \country{Germany}
}
\email{info@sven-mayer.com}

\author{Andreas Butz}
\orcid{0000-0002-9007-9888}
\affiliation{
  \institution{LMU Munich}
  \country{Germany}
}
\email{butz@ifi.lmu.de}

\begin{teaserfigure}
\centering
\includegraphics[width=\columnwidth]{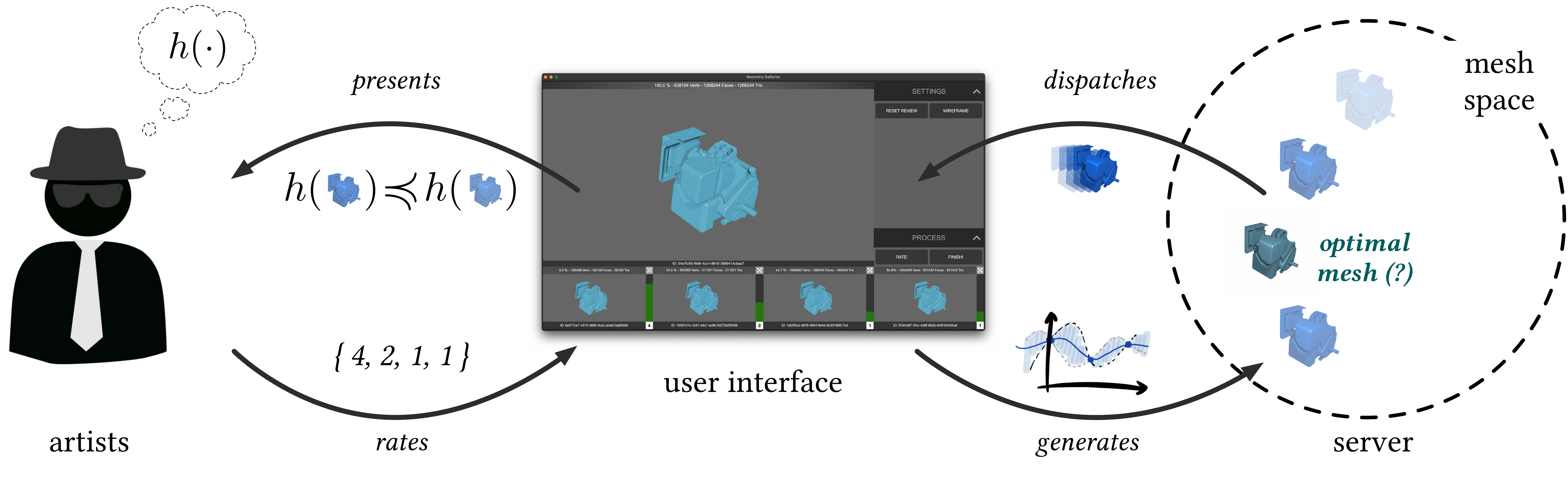}
\caption{
A human-in-the-loop 3D model processing system: A server generates differently processed variations of a complex 3D model and dispatches them to a user interface, which presents those variants to a 3D artist, who in turn rates them. Based on these ratings, new parameter settings are generated and a new set of variations is computed and evaluated again. The process repeats until a satisfactory 3D model is found, that minimizes the number of faces while maintaining as much as possible of its overall appearance.
}\label{fig:diagram}
\end{teaserfigure}

\begin{abstract}
Interactive AI systems increasingly employ a human-in-the-loop strategy. This creates new challenges for the HCI community when designing such systems.
We reveal and investigate some of these challenges in a case study with an industry partner, and developed a prototype human-in-the-loop system for preference-guided 3D model processing. Two 3D artists used it in their daily work for 3 months. We found that the human-AI loop often did not converge towards a satisfactory result and designed a lab study (N=20) to investigate this further.
We analyze interaction data and user feedback through the lens of theories of human judgment to explain the observed human-in-the-loop failures with two key insights: 1)~optimization using preferential choices lacks mechanisms to deal with inconsistent and contradictory human judgments; 2)~machine outcomes, in turn, influence future user inputs via heuristic biases and loss aversion.
To mitigate these problems, we propose descriptive UI design guidelines.
Our case study draws attention to challenging and practically relevant imperfections in human-AI loops that need to be considered when designing human-in-the-loop systems.
\end{abstract}



\maketitle
\renewcommand{\shortauthors}{Ou et al.}

\section{Introduction}
With the increasing interest in human-AI interaction, \emph{human-in-the-loop (HITL)}~\cite{monarch2021hitl} systems have been applied to a wide range of domains, such as material design~\cite{brochu2007activelearn}, animation design~\cite{brochu2010animation_design}, photo color enhancement~\cite{koyam2020seqgallery}, image restoration~\cite{weber2020drawme}, and more~\cite{koyama2014crowd, holzinger2016iml, colella2020interactive_opt, yijun2021melody}.
These systems actively exploit human choices for optimizing machine results. They propose a set of design alternatives and then iteratively adapt their results based on user preference feedback, thereby increasing the quality of the system outcomes and the satisfaction of the human involved while simultaneously speeding up the process.
The ancestor of these approaches is the Design Galleries approach~\cite{marks1997design_galleries}. Its authors state that ``Design Gallery interfaces are a useful tool for many computer graphics applications that require tuning parameters to achieve desired effects''. They proposed a generic user interface (UI) and emphasized techniques for dispersing parameter settings, but they implicitly assumed that the process will always converge and end in a desired solution.

Despite these successful examples, obtaining a desirable result in an interactive process remains a challenging and important issue in general, since it is hard to ``guarantee'', but crucial for such systems to be useful in practice.
We argue that to make progress here, it is particularly insightful for HCI researchers to investigate practical deployments in real workflows of domain experts.
Concretely, our work is guided by the following two research questions, the first of which first relates to our overarching research interest and the second one to the specific case study reported here:
\begin{itemize}
\item[\textbf{RQ1}] \textit{As a domain-specific example, how can the HITL strategy support users in 3D model processing tasks?}
\item[\textbf{RQ2}] \textit{Which challenges arise for human-AI loops in practice and how can we address them?
}
\end{itemize}
To address these questions, we conducted a case study with an industry partner, in which we applied the HITL strategy to the challenging and practically relevant use case of a 3D modeling workflow: Our prototype system helps 3D artists with the problem of \emph{polygon reduction}~\cite{garland1997quadrics, garland1998quadrics} that involves tuning control parameters to remove polygons (mainly triangles or quadrilaterals) from 3D models while preserving their overall appearance (see~\autoref{fig:viscorr}). Although fully algorithmic solutions have improved over the past decades (for instance, using directional fields~\cite{jakob2015instant}), they still suffer from problems~\cite{arvo2015survey} and often produce unusable results for large models or geometric edge cases.
In the meantime, since it often takes 3D artists up to months or years to understand and thus productively work with a new algorithm, this presents a promising target to use an interactive approach in which designers only have to provide their interactive choices to obtain the desired outcome. Thus, we developed a system that optimizes polygon reduction with regard to artists' preferences over the quality of the results based on \emph{Bayesian optimization (BO)}~\cite{shahriari2016boreview, gonzalez2017preferential}. \autoref{fig:diagram} illustrates its general architecture.

After fine-tuning the system using pilot usage feedback, two professional 3D artists used our system for three months in their real-world 3D workflow. Analyzing this deployment and its usage logs, we found a high failure rate when using a human in the loop. Crucially, final ratings from artists diverged from partially inconsistent or inexplicable compared to their previous ratings.
This mismatch motivated us to conduct a follow-up study (N=20) to confirm these failures in a controlled setting. We analyze interaction data and user feedback through the theories of human judgment to explain the observed failures of the human-AI loop with two key insights: 1) optimization using preference lacks mechanisms to deal with inconsistent and contradictory human judgments; 2) machine outcomes, in turn, influence future user choices via heuristic biases and loss aversion. In this light, we conclude the paper by proposing descriptive UI design guidelines for mitigating these practical problems.
\section{Related Work}
We will first briefly summarize recent work in BO on preference learning and polygon reduction methods in geometry processing, which drove our system implementation. Then, we will discuss the fact that recent HCI research emphasized the benefits of using HITL to support user or algorithm efficiency, but that, on the other hand, little is known about what might keep such a system from working properly.

\subsection{Learning Human Preferences using Bayesian Optimization}
As processing human subjective input is challenging, modeling human choices~\cite{fuernkranz2003pref, chu2005preference} is a research topic in preference learning. To exploit preferences, formal models consider a human rater as an unknown utility function that produces ratings, such as on a 5-star scale, concerning evaluation instances.

The optimization of control parameters based on feedback from a human relying on domain knowledge can be described as an optimization problem: Search an optimal \emph{parameter set} $p^*\in \mathcal{P}$ such that $p^* = \operatorname*{argmax}_{p\in \mathcal{P}} h(M(p))$ where $p$ is any parameter set in the \emph{parameter space} $\mathcal{P}$, $M(p)$ is the instance generated using parameter set $p$, and the \emph{preferential function} $h$ returns the human rating of system output $M(p)$.

When querying the function $h$ is costly, for example, because it involves asking a human for a preferential decision, BO is often selected~\cite{brochu2007activelearn, brochu2010animation_design, koyam2020seqgallery, mikkola2020projective} and has provided a wide range of successful applications~\cite{shahriari2016boreview, holzinger2016iml, koyam2020seqgallery, koyama2021ui}. BO actively learns a \emph{posterior} from human preference based on the collected ratings and can then predict an estimated optimal $p^*$ in the next iteration step. Thus, this method captures how artists usually develop their expertise with a new system by exploration and exploitation, resulting in an overall improvement after some iterations.

As a variant, preferential Bayesian optimization (PBO)~\cite{gonzalez2017preferential, koyam2020seqgallery, mikkola2020projective} is an improved version of BO. It evaluates preference on paired samples with an adaptive reference point. The reason for using PBO is that humans are better at differentiating paired samples than they are at determining absolute values~\cite{chu2005preference, brochu2007material_design, brochu2007activelearn} according to Thurstone's law of comparative judgment~\cite{thurstone1927comparelaw}. Therefore, it is considered better for use in HITL systems, and we integrate PBO to learn human preferences.

\subsection{Polygon Reduction in Geometry Processing}

For the specific geometry processing problem, polygon reduction, presented in this paper, we briefly discuss its recent advances and the relevant methods we utilized to build our system. The geometry processing ``No-Free-Lunch'' theorem~\cite{wardetzky2007no_free_lunch} states that not all geometric properties can be well preserved simultaneously in discrete instantiations of a smooth geometry. Therefore, different processing tasks have specialized algorithms and corresponding configurations, such as soft or hard geometries.
In general, polygon reduction methods can be categorized as \emph{local decimation}, \emph{global remeshing}, or a weighted combination of both.

Local decimation means that neighbor vertices and edges are greedily removed. These methods date back to the last century~\cite{garland1997quadrics, garland1998quadrics} and have also been used for levels of detail (LOD) generation~\cite{hoppe1996progress}, even baked into hardware rendering pipelines~\cite{olano2003lodshader}. They are efficient but contain ill-posed cases with results depending on the implementation.
Instead, the general idea of global remeshing~\cite{bommes2013quadsurvey, knoppel2013globalfields, ebke2014lodquad, pietroni2021featureline} is to define a \emph{directional field} as constraint boundary conditions on a \emph{Poisson equation}, and then minimize an artificial energy function. After minimization, a new target mesh can be reconstructed from scratch using the solved solution. Computationally, this is much more costly and cumbersome, but the resulting mesh quality is much better than that from local decimation.
State-of-the-art practical solutions, such as \citet{karis2021nanite}, use a weighted combination of both that balances the processing speed against quality. A large mesh can be split into smaller ones, then processed using mixed global~\cite{jakob2015instant, huang2018quadricflow} and local~\cite{hoppe1996progress, garland1998quadrics} methods, but this also introduces the new problem of cutting a mesh. We refer to Metis~\cite{karypis1997metis, ponchio2009graphcut} as a mature solution for graph partitioning.

\subsection{Decision Error in Human Judgments}
\label{sec:decision-error}
Economics widely studied decision-making when choosing a preferred item from several alternatives. The expected choice utility maximization~\cite{marschak1960choiceutility} forms the theoretical basis. It describes a standard economic model on a finite number of decisions but assumes that individual beings behave rationally. In psychology, \citet{simon1955rational} proposes the concept of \emph{bounded rationality} and proposes to replace this assumption, as rationality is only limited, and  decisions are made by \emph{satisficing}. Later, \emph{prospect theory}~\cite{kahneman1979prospect, tversky1992advprospect} empirically demonstrated human judgments in reality when trying to maximize a certain utility function (wealth) in risky situations (e.g., under time pressure) and explained the behavior with bounded rationality.
The \emph{heuristic biases} constitute a key source of general decision error.
\citet{kahneman1974juedgment} showed that in any decision under uncertainty, System~1 (fast and instinctive thinking) tends to override System~2 (slow and rational reasoning)~\cite{kahneman2011thinking}, hence creating a statistical bias on the decision. More specifically, 1)~\emph{representativeness} substitutes the most readily accessible examples to form a decision, 2)~\emph{availability} uses mental shortcuts, and 3)~\emph{anchoring} as a conclusion bias describes that initial information has a consequence on a later decision.

In addition to heuristics, other effects can also influence judgments: 1)~in a utility maximization context, \emph{diminishing returns}~\cite{shephard1974diminishing} may occur as wealth increases and marginal utility decreases; 2)~\emph{loss aversion}~\cite{kahneman1972subjectiveprob}, as part of the \emph{endowment effect}~\cite{kahneman1991endowment}, describes that people prefer to retain an owned property rather than to acquire an alternative, potentially better one. People hence tend to stick to seemingly safe decisions when a potential gain would require more risk.
3)~In the present understanding, combined with a statistical view~\cite{brandstaetter2006tradeoff,gigerenzer2009homo}, systematic noise descriptively shapes another form of decision error that contributes equally to judgment error as individual bias~\cite{kahneman2021noise}. The decision noise components~\cite{kahneman2021noise} break down into \emph{level noise} (decision variability between groups), \emph{stable pattern noise} (contextual bias within groups), and \emph{transient noise} (purely occasional).

\subsection{Human-in-the-Loop Systems}
The HITL strategy may be applied in different contexts, which connect to different fields, including personalization, co-creation, and decision-making support.

From the human computation~\cite{vonahn2008human, quinn2011human} perspective, a HITL system may be designed to use crowds~\cite{ipeirotis2010quality} as human processors to solve system tasks that neither machine nor human can solve independently. Although using collective intelligence has been largely verified to be beneficial for crowdsourcing tasks~\cite{kamar2012combining}, there are several identified challenges~\cite{reeves2010five} to integrating human computation, which highlighted challenges such as user motivation, sustainability, and input bias. To motivate users to contribute, researchers have used a Game-With-A-Purpose (GWAP) approach~\cite{krause2011human}, but turning a task into a game could be another challenging design problem. Dealing with diverging opinions within small crowds may be difficult because tasks might require a certain level of expertise. Moreover, HITL systems using crowds may suffer from malicious inputs~\cite{ou2019gwap} and lead the entire system toward using biased inputs when the initial samples lack trust.
Particularly for design-related tasks, crowd opinions may not fit individual interests and needs regardless of data bias. Hence using crowd-powered design systems~\cite{koyama2014crowd} is considered limited when individual customization has a higher priority.

In a personalized context, \citet{buschek2021hai_pitfalls} examined the potential pitfalls for achieving user interests in the co-creation context. The limitations on the machine side, identified as lack of machine \emph{creativity}~\cite{lamb2018creativity}, and \emph{usability}~\cite{kocielnik2019imperfectai}, and thus, highlight a biased AI with trained system bias but lacks discussions on the source of bias and the mismatch between individual expectations and system abilities.
In terms of mismatched expectations, \citet{eiband2019meet} reported that users might intentionally provide flawed inputs when a system fails to achieve their satisfaction in everyday intelligent applications. As a follow-up, however, \citet{voelke2020trickai} showed that a user must exhaustively provide noisy feedback to confuse an intelligent system. Still, they lack verification and interpretation as to whether the repeated unsatisfactory results come from system limitations or human behavior change.
For AI-assisted decision-making scenarios, \emph{trustworthiness} becomes a primary social concern regarding \emph{reliability} in areas where a decision is vital, such as clinical decisions~\cite{cai2019medical, knapic2021medical}. Still, it is \emph{implicitly assumed} that the human involved eventually makes a rational decision over \emph{subjectively} untrusted AI outcomes. Factors such as algorithm aversion~\cite{dietvorst2015algorithm_aversion} were confirmed to indicate that users are more biased~\cite{ragot2020perception_bias} towards human results and produce considerable noise even in the judicial area~\cite{dressel2018recidivism}.

Although prior research~\cite{marks1997design_galleries, koyama2014crowd, holzinger2016iml, weber2020drawme, yuksel2020hitlml, colella2020interactive_opt} that involves HITL strategies have shown human knowledge to be helpful for a machine to learn, previous literature rarely discusses the circumstances under which HITL could shine. Especially when users intentionally or unintentionally provide defective or uncertain inputs, it is unclear whether the system can continue to process it effectively and whether other cascading effects will be triggered.
Our paper addresses this research gap and shows challenges that can arise when exploiting individual preferences in a HITL system in practice.
\section{System Design and User Workflow}
\label{sec:system}
\begin{figure*}
\centering
\includegraphics[width=\textwidth]{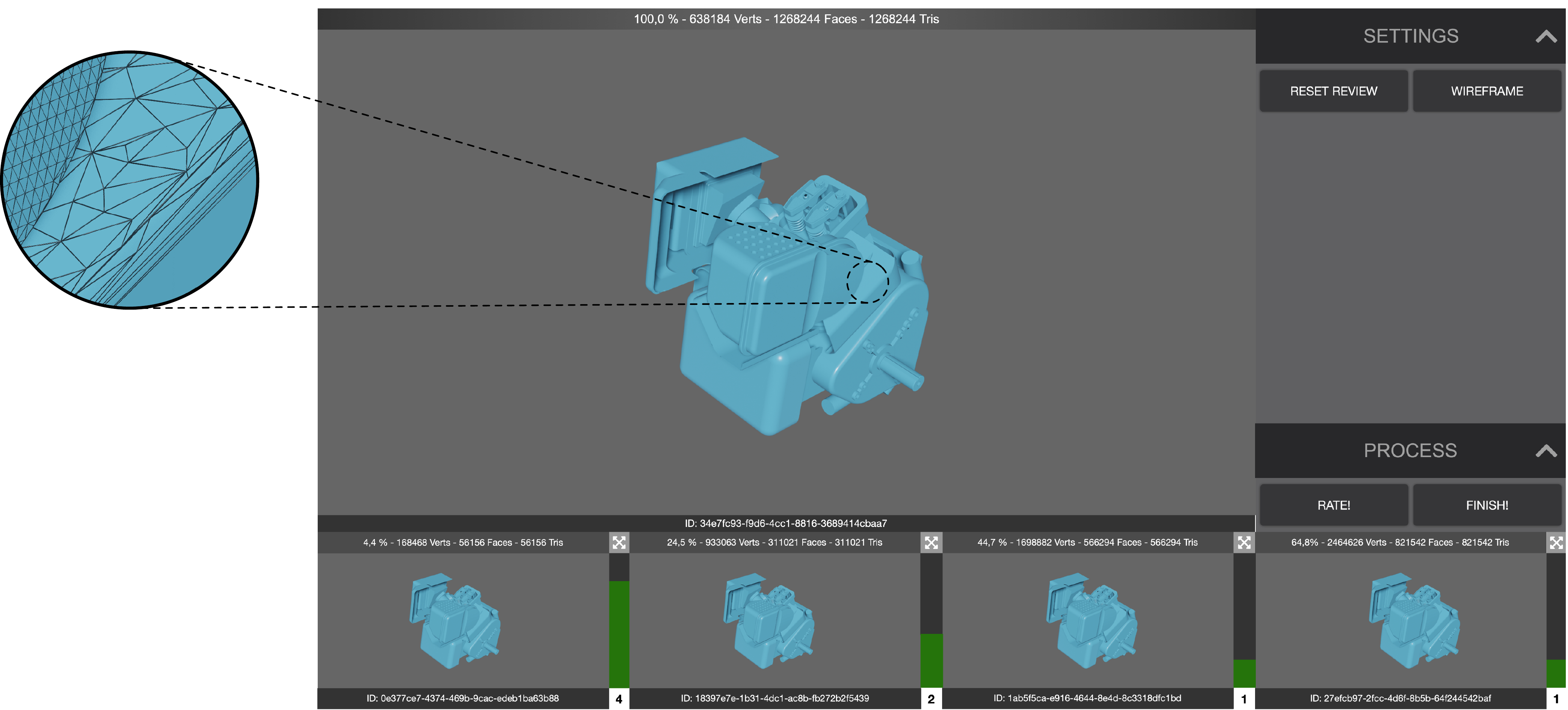}
\caption{The UI for users to rate variant models. An artist can move each mesh variant to the bigger, central view and activate a wireframe for detailed quality inspection, which was considered necessary in previous work~\cite{botsch2010polygon} because of the alignment of mesh edges is considered a quality metric. Artists must make a professional choice between visual rendering quality and wireframe quality, especially in cases that may sacrifice a bit of wireframe quality for substantial gains in polygon reduction (lower number of triangles).}\label{fig:interface}
\end{figure*}

We approached the polygon reduction problem in a joint project with an industrial collaboration partner and were motivated by the goal to optimize their artists' daily tedious manual tuning workflow. One of the benefits of having simplified 3D mesh models is to reduce rendering complexity due to fewer primitives required in a render pass. 3D mesh-based models in the project ranged from a single water-tight mesh layer to models that were either manually sculpted or automatically converted from solid geometry data formats, containing hundreds of sub-meshes with millions of polygons in total. Domain experts processing these mesh models usually require days or weeks of extensive manual work.

We expose the core functionalities of our system as Web APIs that run the polygon reduction on a GPU server. The developed frontend UI by our industrial partner uses Unity\footnote{\url{https://docs.unity3d.com/2020.3/Documentation/Manual/UIToolkits.html}} as shown in \autoref{fig:interface}.
We engineered a hybrid local/global algorithm based on state-of-the-art research~\cite{garland1998quadrics, jakob2015instant} controlled by nine different parameters for the polygon reduction itself. We determined the initial parameters by domain heuristics in a few pilot tests with our industrial partner.

In our designed workflow, an artist can first upload an original model. The server then simplifies the uploaded model under different parameter settings in the background. When it has computed all alternatives, taking between seconds and minutes, they are downloaded back into the UI. When the artist indicates their ratings of model quality, the system learns from these judgments and continues the process again to generate more optimized models. The rating scale for judgments is 0~(\emph{skip}, meaning not considered due to faulty geometry), 1~(\emph{terrible}) to 5~(\emph{excellent}). Without loss of generality, if the human decided ratings for the four variant models $M_{i} (i = 1, 2, 3, 4)$ are 3, 4, 5, and 1, then this represents six preferential choice relations: $M_1 \preccurlyeq M_2$, $M_1 \preccurlyeq M_3$, $M_2 \preccurlyeq M_3$, $M_4 \preccurlyeq M_1$, $M_4 \preccurlyeq M_2$, and $M_4 \preccurlyeq M_3$ where $\preccurlyeq$ means ``is less preferred.'' 3D models in the next iteration are optimized based on these relations using PBO which made us expect~\cite{gonzalez2017preferential, mikkola2020projective} the system to converge to the desired outcome quickly.
Note that the model quality is \emph{not} equivalent to just visual rendering quality but also involves edge flow and other geometric properties~\cite{botsch2010polygon} that require a subjective decision. Hence human judgments in this task involve visual rendering quality, mesh wireframe alignment quality, and other technical metrics such as volume-preserving, which is also why this task is more preferred to query human opinions than done by full automation.

\section{User Studies}
To evaluate the effectiveness of our system, we conducted two studies; first, a field study with our industrial partner, and second a lab study to verify the effects of the field study. In the field study, we used our system in a real-world setup with our industrial partner, who aimed to optimize their process in customer projects. Here, we had the opportunity to run a field study where designers used the system in their daily workflow. However, due to the uncontrolled environment of field study, making in-depth assessments and conclusions on specific aspects can be hard. Thus, we additionally ran a lab study to understand the failure cases of our system to verify our findings further under controlled conditions.

\subsection{Field Study}
For the field study, we first conducted a set of small pilot experiments to fine-tune our system's parameters to fit the partner's needs and their customer projects.

\paragraph{Participants}
We recruited two full-time 3D technical artists from our industry partner to gain insights into the newly developed workflow involving our interface. Both are male, aged 25 and 35, one has more than three years of experience, and the other has more than eight years of experience in the 3D industry.

\paragraph{Procedure}
The two experts used our system almost daily to evaluate model quality during polygon reduction.
However, they were not restricted to our interface and could also use further software aids (as in their previous workflow) for the model quality inspection, e.g., for accessing more professional curvature visualizations.
When using our tool, the loaded 3D model was computed into four variants. After the experts finished their evaluation (either inside our interface or externally), they rated the models in our interface. These ratings were used for the next optimization iteration to generate new variants (see \autoref{fig:diagram}). The rating process terminated when the experts found the results satisfactory or reset it.

\paragraph{Collected Dataset}
During the three months of the study, we collected 549 evaluation sequences as a field study dataset. This corresponds to 4.5 evaluations per expert and workday.
Of these, 415 sequences terminated in the first iteration without any preference optimization requested. The remaining 134 sequences (number of iterations: $\mu=4.1, \sigma=4.2$, range 1-23) contain sequential preference ratings.
The collected 3D models included various mesh types, including organic, soft surfaces, hard technical surfaces with sharper angles, and combinations, such as machinery parts (see~\autoref{fig:interface}) with both smooth, flowing lines and hard, mechanical edges.

\subsection{Lab Study}
As a lab study, we conducted a within-subjects user study to further understand our system's use with specific assistance information and behaviors in a larger user group with different backgrounds in 3D.

\paragraph{Participants}
We recruited 20 participants using convenience sampling (7 female and 13 male; age $\mu=27.0, \sigma=8.8$, range 18-62). Among them, four had more than one year of industrial experience in 3D modeling, and all others had no experience.

\paragraph{Procedure}
\begin{figure*}%[ht]
\centering
\includegraphics[width=\textwidth]{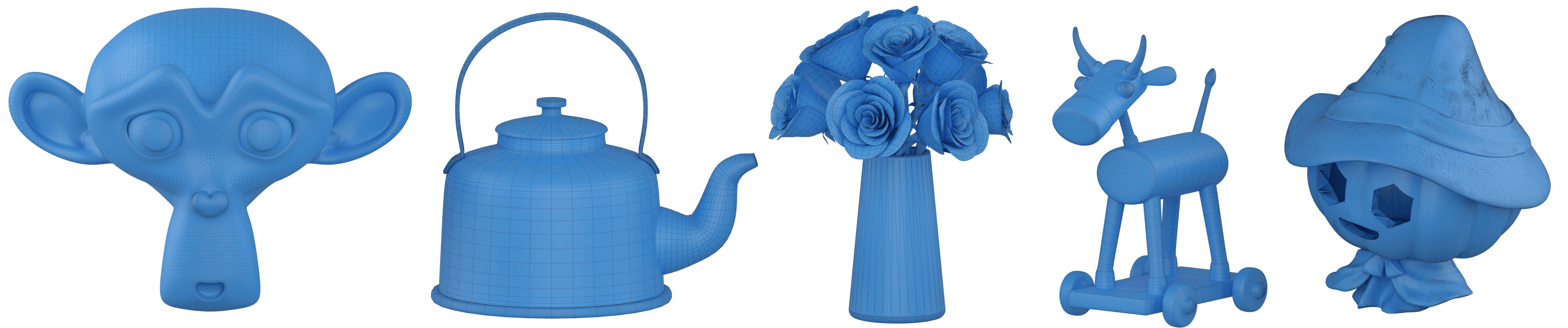}
\caption{Models that are selected in the lab study, from left to right: \emph{monkey, teapot, rose, cow, pumpkin}. Each model was rendered \emph{with} and \emph{without} wireframes separately.}
\label{fig:lab-models}
\end{figure*}

We first welcomed participants and explained the study, answering all open questions before they signed the consent form. Then, participants were presented with different 3D objects in every evaluation session.
The overall procedure in terms of rating and termination process in each evaluation session was similar to in the field study. In detail, we asked participants to balance the trade-off between polygon reduction and quality loss. Thus, they had to indicate their preference using ratings to optimize models iteratively. Participants could terminate fast in a few iterations if satisfied or were stopped at the 11th iteration to limit the time of participation. After each evaluation session, we asked participants 7 questions (see \autoref{sec:open-source}) to indicate their satisfaction and provide overall subjective feedback.
We selected five different 3D models, as shown in \autoref{fig:lab-models}, and each model was rendered \emph{with} and \emph{without} wireframes (instead of allowing users to activate them freely). We picked these five models with the two wireframe representations to ensure our results generalize beyond this small set of objects. We displayed the order of these 3D models and their wireframe display using a Latin square design to avoid learning and fatigue effects. Therefore, we collected $5\times 2=10$ evaluation sessions in total for each participant. On average, each participant spent 90 minutes in the entire study.

\paragraph{Collected Dataset}
We collected 200 evaluation sequences (number of iterations: $\mu=5.1, \sigma=2.9$, range 1-11) by design, and all sequences involved at least one preference optimization.
The selected model covers a similar spectrum of models as the experts had experienced in our field study. These models were also simpler than complex real-world models to reduce the time of machine optimization and participation waiting time.

\section{Results}
\label{sec:results}

Below, we report our analysis of the participants' rating process using collected data from the two studies, also in comparison to each other, and show that if participants are not rated by pure random, they at least behave highly unstable and inconsistent in the rating process. Then, we show selected example cases from the collected data that were also discussed with experts in hindsight concerning why they made a particular rating choice.

\subsection{Human-AI Mutual Interventions}

\paragraph{Overview}
In the \emph{field study}, from the 134 sequences with preferential ratings, only 16 sequences ($11.9\%=16/134$) produced a satisfactory outcome. In the \emph{lab study}, among the collected 200 sequences, only 97 sequences ($48.5\%$) were terminated with a satisfactory outcome. Both studies suggest a high failure rate in optimizing HITL outcomes.

\begin{figure*}[t]
\centering
\begin{subfigure}[b]{0.49\textwidth}
    \centering
    \includegraphics[width=0.925\textwidth]{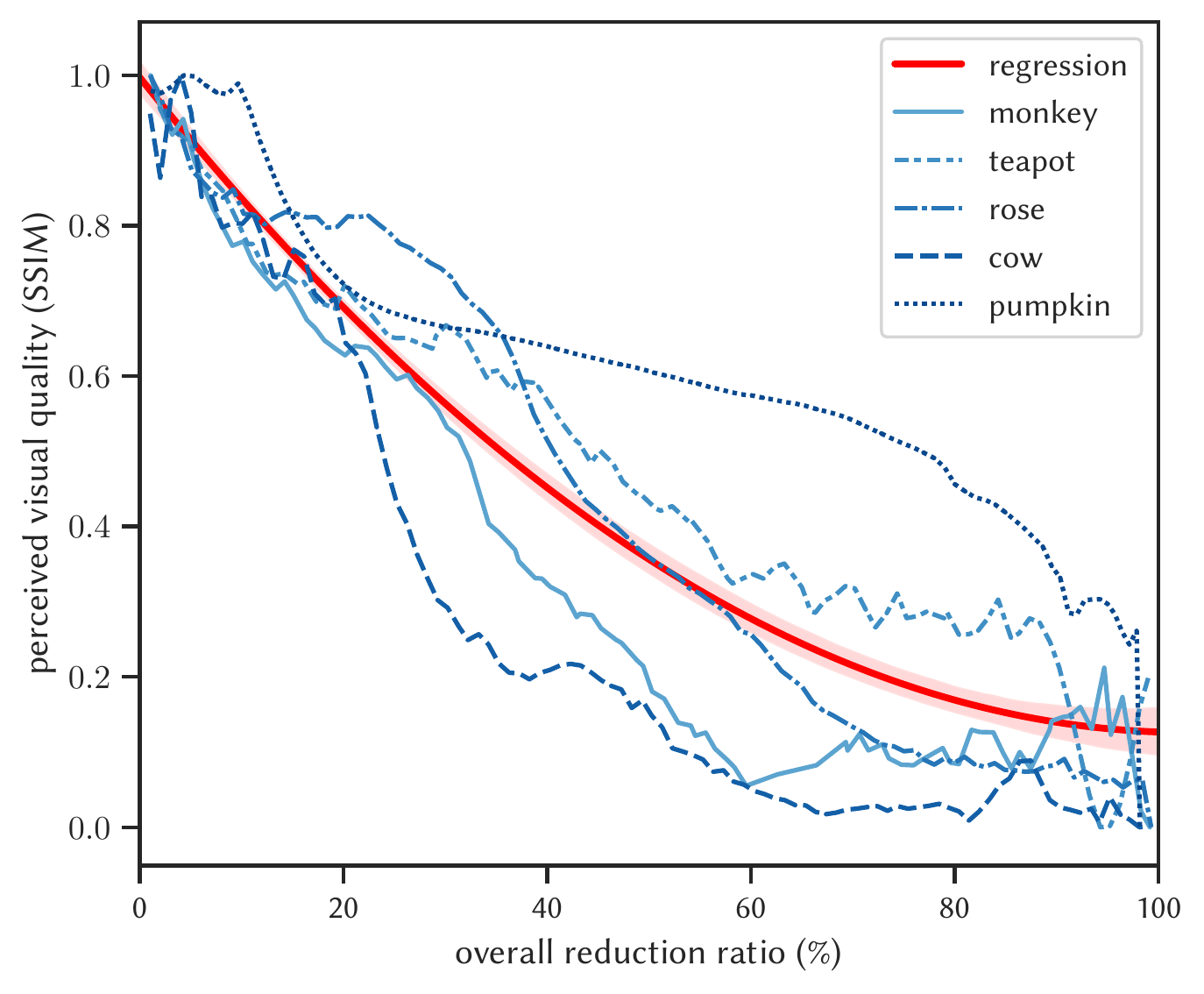}
    \caption{The relation between reduction ratio of models in lab study and human perceived visual quality (SSIM, normalized), and the red solid smooth curve shows a second-order polynomial regression.}
    \label{fig:viscorr}
\end{subfigure}
\hfill
\begin{subfigure}[b]{0.49\textwidth}
    \centering
    \includegraphics[width=0.925\textwidth]{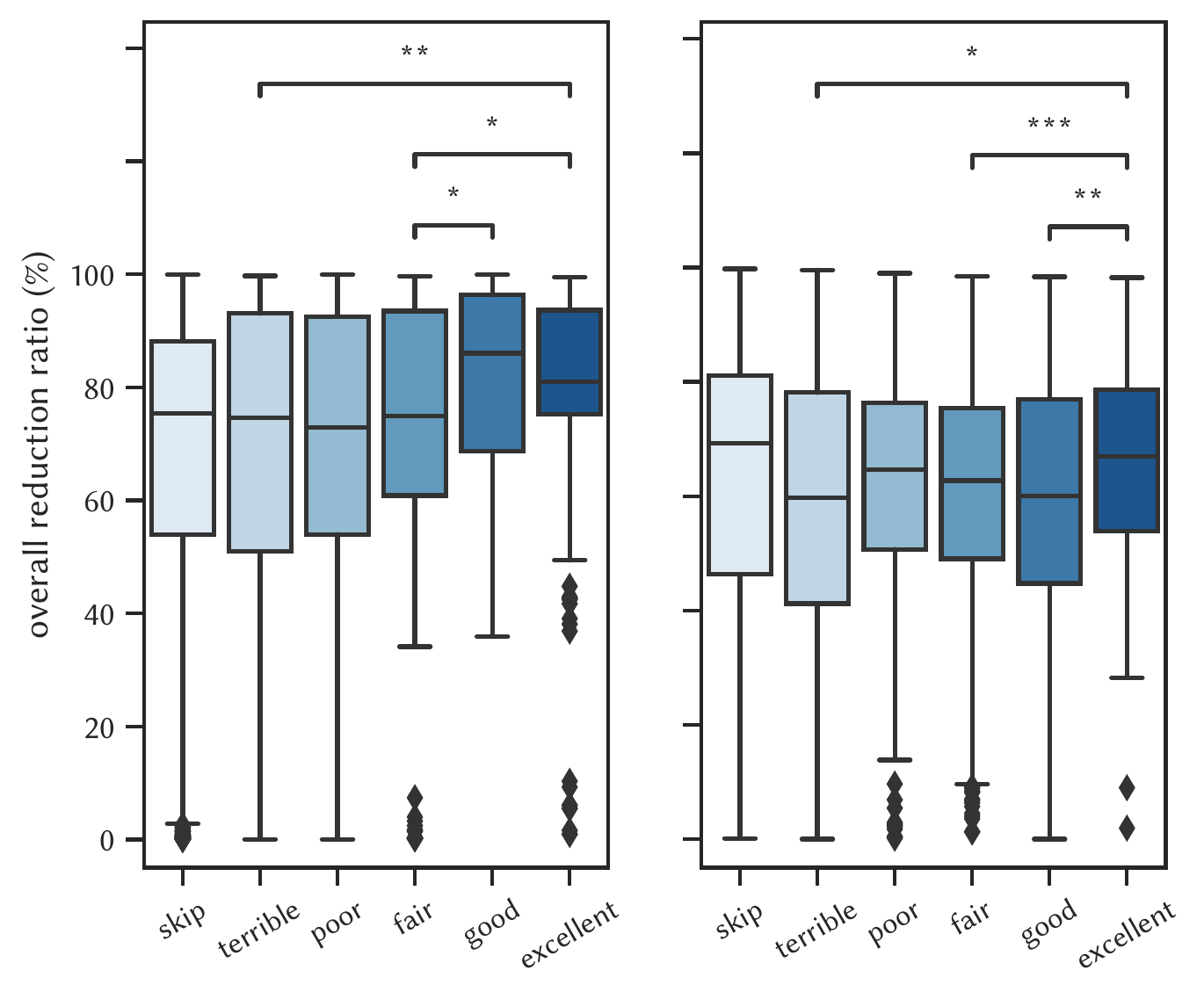}
    \caption{Field (left) and lab (right) studies' preferential ratings against overall reduction ratio (higher means stronger reduction), *~($p<.05$), ** ($p<.01$), *** ($p<.001$).}
    \label{fig:ratings}
\end{subfigure}
\caption{Overview of the visual influences of polygon reduction and collected rating data.}
\end{figure*}

\paragraph{Effectiveness of Human Judgments}
\autoref{fig:viscorr} shows a second-order polynomial regression between a human perceived quality of 3D models used in our lab study and the overall reduced amount of polygons. We assess perceived visual quality using an average of multiple \emph{structure simularity} (SSIM)~\cite{wang2004image} that compares the rendered visual quality between the reduced and original 3D model in five different camera views. The \emph{reduction ratio} represents the removed polygon count of a resulting model divided by the total polygon count in the original model.
\autoref{fig:ratings} shows the rating distributions in the two studies regarding reduction ratio.

We used Kendall's $\tau$ coefficient to measure the ordinal association between the reduction ratio and rating scale. The result shows a significant correlation ($\tau=0.07, p<.001$) in the field study whereas no significance ($\tau=0.004, p=0.71$) in the lab study. This suggests that field study experts tend to give higher ratings to highly reduced models, but lab study is more diverse.
For a more fine-grained measure between rating scales, we also used Mann–Whitney U tests to check for dependencies between different rating scales and the reduction ratio:
1)~We found an significant difference between \emph{fair} ($Mdn=74.88$) and \emph{excellent} ($Mdn=80.95$) ratings~($U=8756.0$, $p<.001$) and a significant difference between \emph{terrible} ($Mdn=74.65$) and \emph{excellent} ($Mdn=80.95$) ratings~($U=10219.0$, $p=.003$), i.e. in cases where reduction ratio was positively correlated with rating.
On the other hand, we found no significant differences in reduction ratio between \emph{terrible} ($Mdn=74.65$) and \emph{poor} ($Mdn=72.93$) ratings~($U=19705.0$, $p=.62$) or \emph{good} ($Mdn=80.03$) and \emph{excellent} ($Mdn=80.95$) ratings~($U=3602.0$, $p=.37$), i.e., where there would have been a negative correlation. \emph{In sum, this suggests that the collected ratings are effective to the highly reduced models, and the reduction ratio is one of the effectively relevant factors in human judgments}.
2)~We found no significant differences in highly reduced models between good and excellent in the field study, but a significant difference between \emph{good} ($Mdn=60.03$) and \emph{excellent} ($Mdn=66.99$) ratings ($U=131031.0, p=.002$) in the lab study. Although the field study had fewer users, this could also be interpreted such that \emph{experts in the wild use other quality metrics, which lab participants with less expertise overlook}.
3)~Models rated as \emph{good} and \emph{excellent} have higher mean reduction ratio in the field study ($M_{\text{good}}=81.68, M_{\text{excellent}}=75.52$) than in the lab ($M_{\text{good}}=61.10, M_{\text{excellent}}=64.84$, also see~\autoref{fig:ratings}), which suggests that \emph{lab study participants are easier to satisfy by the system outcomes than expert artists}.

\begin{figure*}[t!]
\centering
\includegraphics[width=\textwidth]{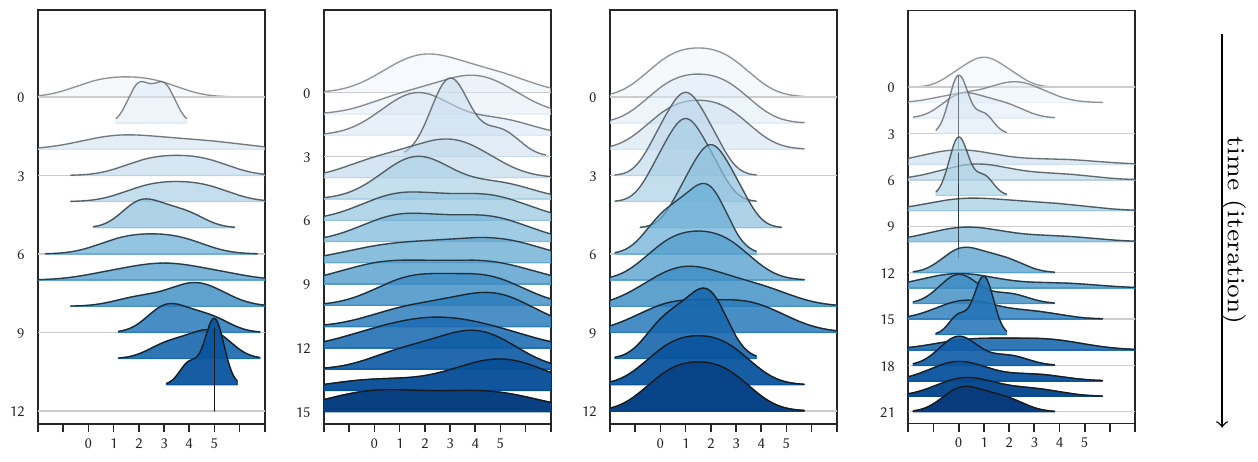}
% \includegraphics[width=0.23\textwidth]{assets/ideal2.pdf}
% \includegraphics[width=0.23\textwidth]{assets/dacdd183-4dd0-11ec-86eb-a85e4557a9b6.pdf}
% \includegraphics[width=0.23\textwidth]{assets/48f9c537-43f4-4a22-a7ff-4ce8d9e48d4e.pdf}
% \begin{tikzpicture}
%     \pgftext{
%     \includegraphics[width=0.23\textwidth]{assets/086d8b7a-fa86-4b80-93b7-1a40fed9bc80.pdf}
%     }
%     \draw[<-] (2.18,-2) -- (2.18,2);
%     \node[rotate=-90] at (2.3,0) {\scriptsize time (iteration)};
% \end{tikzpicture}
\caption{Comparing an ideal (far left) and actual (the others) rating distributions over time (from top to bottom), the bottom axis 0 to 5 represents the rating scale. An expected rating distribution should move to the right side over time if users are more satisfied with the results, but the actual preferences drift back and forth between 0~(\emph{skip}) and 5~(\emph{excellent}).
}
\label{fig:rating-dist}
\end{figure*}

\paragraph{Stationarity and Trends of Data}
\autoref{fig:rating-dist} compares how an ideal (far left) and three actual rating distributions (the rest) drift over time: In our context, since the objective of using PBO is to search a polygon reduction configuration to maximize the human ratings~\cite{gonzalez2017preferential}, ideally, in a successful exploring and exploiting sequence of preferences, and the mean rating score should \emph{increase} and drift from low values with high variance towards higher values with lower variance (\emph{non-stationary} and with an increasing \emph{linear trend component}).
However, the actual sequence shown (as most others) stagnates and fluctuates back and forth. From the 200 sequences collected in the lab, 79 continued to at least four iterations (required for the subsequent trend test), and we tested them using an Augmented Dickey-Fuller test. Results show that 36 rating sequences are stationary ($p<.05$). In the remaining 43 non-stationary sequences, a Mann-Kendall test found only four significant increasing trends in the mean rating score ($p<.001$) and only one significant decreasing trend of rating variance ($p<.05$). Another Mann-Kendall test found that only three sequences had increasing and six sequences decreasing trends regarding the machine-estimated optimal reduction ratio ($p<.05$).

In summary, all these results imply for the optimized process, that
1)~on the human side, the rating behavior does not improve over iterations;
2)~on the machine side, the optimized reduction ratio using preferential choices does not improve over iterations.
This suggests that \emph{the human-machine loop as a whole is kept from terminating and fails}.

\begin{figure*}[t]
\centering
\begin{subfigure}[b]{0.49\textwidth}
\centering
\includegraphics[width=\textwidth]{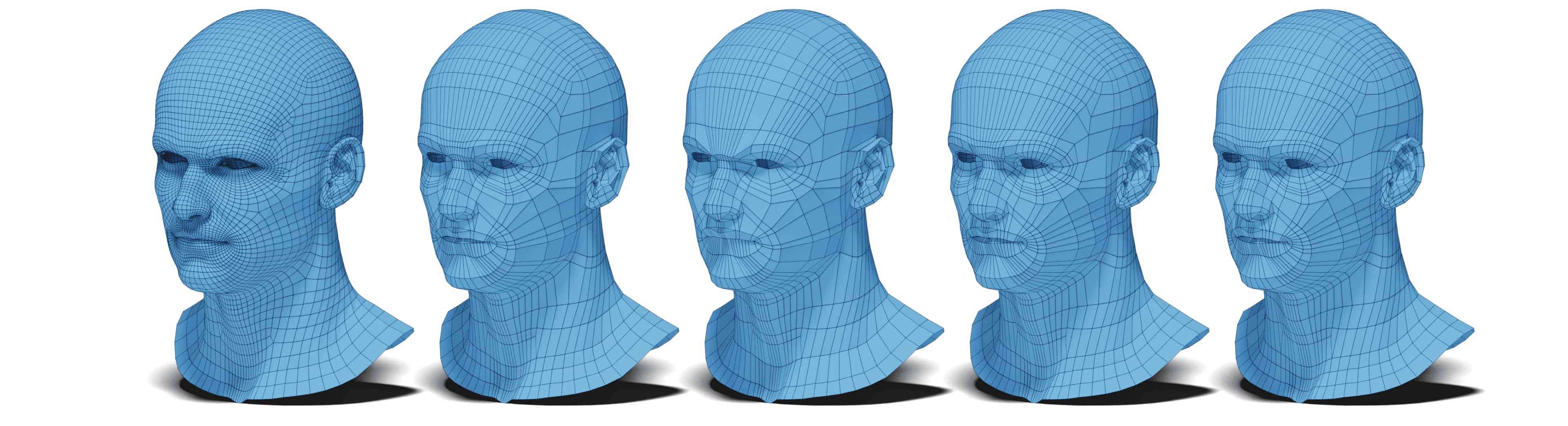}
\includegraphics[width=\textwidth]{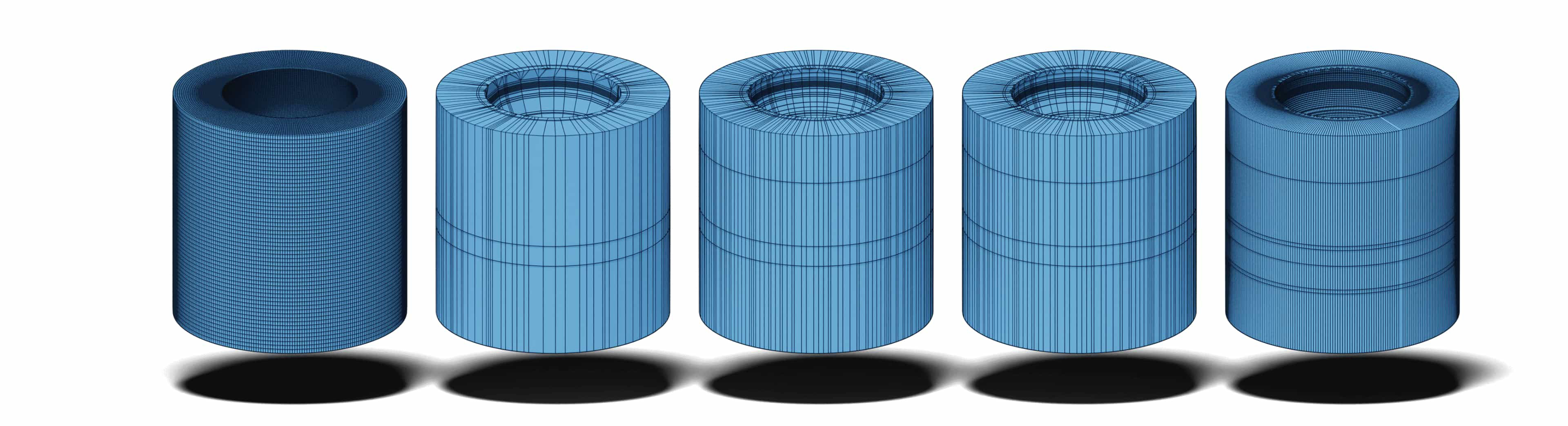}
\caption{
In the same iteration, original (far left) and 4 processed models: 1) Head model: the last 2 models are identical, ratings (left to right): 5, 3, 4, 1; 2) Cylinder model: the middle two are almost identical. Ratings: 5, 4, 3, 2.
}
\label{fig:head-cylinder}
\end{subfigure}
\hfill
\begin{subfigure}[b]{0.49\textwidth}
\centering
\includegraphics[width=\textwidth]{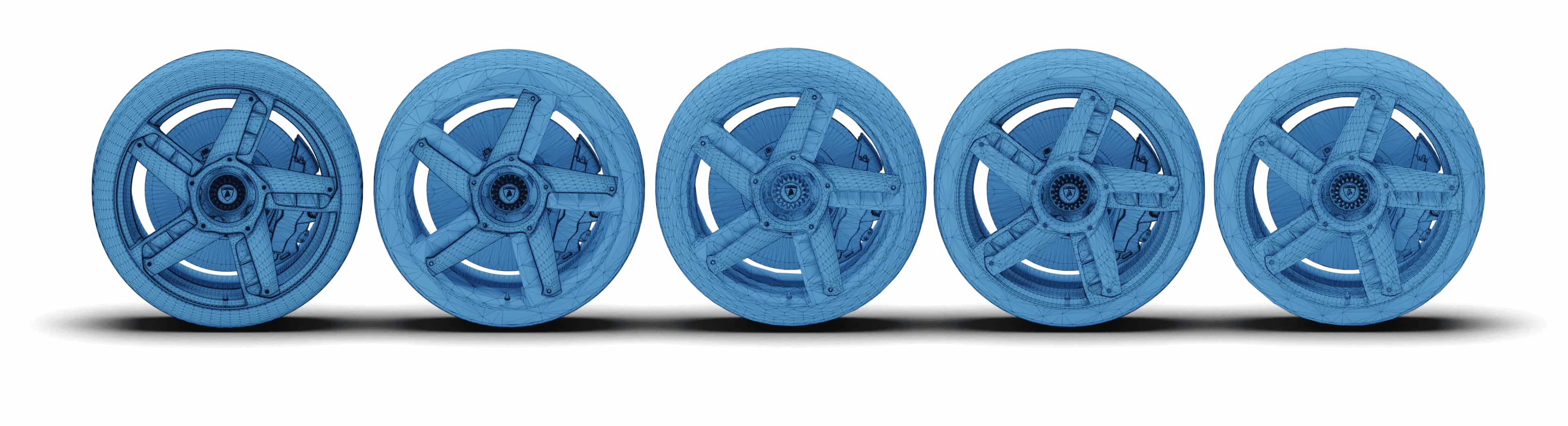}
\includegraphics[width=\textwidth]{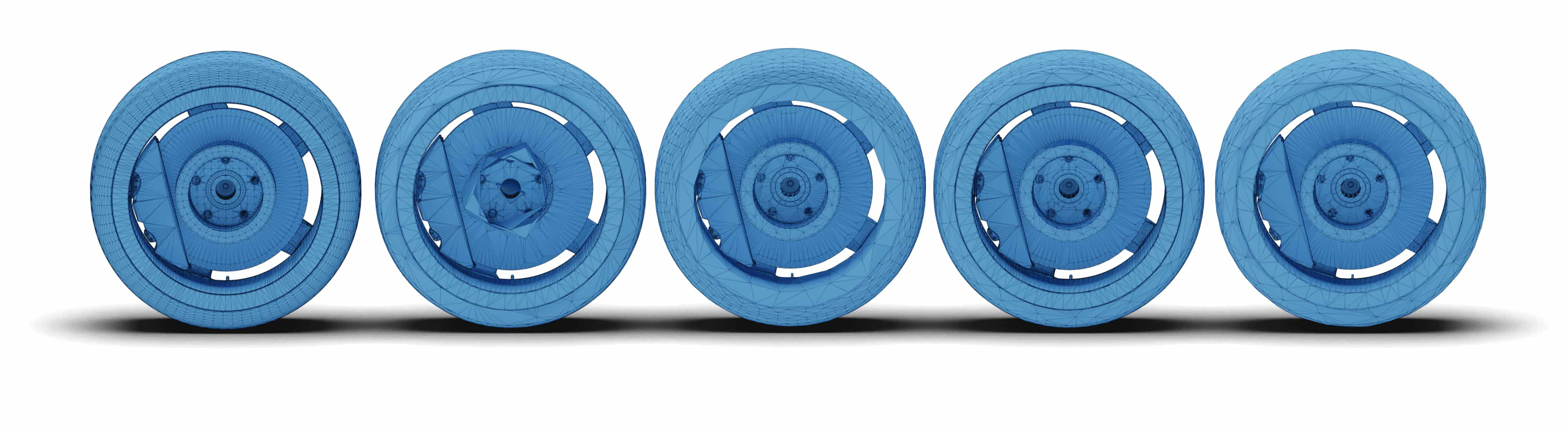}
\caption{
Original (far left) and 4 reduction variants, all rated as 5 in the same iteration. The bottom row shows the backside of the models. Objectively, the middle left wheel contains faulty geometry and should receive a 0 (skip) rather than a 5 (excellent).
}
\label{fig:wheels}
\end{subfigure}
\caption{Selected examples that were discussed in the semi-structured interviews.}
\label{fig:models}
\end{figure*}

\subsection{Semi-structured Interviews with Experts}\label{sec:semi_structured_interviews}
In the semi-structured interviews, we discussed with the two expert artists, case by case, inconsistently judged models and why they made a certain (contradicting) choice. \autoref{fig:models} shows three of the discussed models.
\autoref{fig:head-cylinder} contains a head model and a more straightforward example of a CAD-converted cylinder that is cut by a sphere.
Below, we discuss three example cases in more detail:
Ironically, the two head models on the right are identical but received an entirely different rating scores of 4 and 1 from the same rater in the same round (\emph{\setword{case 1}{case:1}}). The middle right cylinder (rated 3) compared to the middle (rated 4) contains fewer polygons and better symmetries but received a lower rating (\emph{\setword{case 2}{case:2}}).
As a slightly more complex case (\autoref{fig:wheels}), a wheel model with reduced variants that all were rated a 5 in the same iteration, but the middle left one did reduce to faulty geometry on the backside (\emph{\setword{case 3}{case:3}}), which the artist had missed.

In \ref{case:1}, \emph{artist A} confirmed that his ratings strongly depended on what he had seen before and admitted that he tended to give one model a terrible rating in each iteration due to previous rating experience. He also mentioned that ``\emph{...I sometimes stopped giving a higher score because I had decided on a different objective}'' in processing the model (e.g., to go for more visual quality but less reduction).
In \ref{case:2}, \emph{artist B} argued that he had scored the middle cylinder mesh higher than the middle right one because ``..\emph{the usually difficult inner hard edges were handled better}..'' in that case. However, the middle right model has an objectively higher reduction ratio, and both contained similar defects on the inner hard circular edge. Their differences are only at a technical level.
Furthermore, he explained that in \ref{case:3}, he did not notice the flaw at first sight and rated the wheel a 5 simply because it was shown from the front. He had made a quick decision based on the visible mesh quality of the tire and based on a similar experience, which is an example of a reasonably simple oversight with potentially harmful consequences. The artist also explained that after many iterations, ``\emph{...it gets frustrating to see the more flawed output after I already had seen a partially good result}.''
\section{Discussion}
Although our evaluation does not examine any entangled causality but only statistical correlation, it is likely that the observed system failure initially starts from human error as the system was initialized with the same prior in each of the sequential evaluations (using a Matérn kernel ($\rho=2, \nu=2.5$) with the statistical properties of being isotropic and stationary~\cite{rasmussen2004gaussian, shahriari2016boreview}). Since errors are further propagated and amplified to system outcomes, we combine theories regarding human decision errors to reflect on and explain our findings.

\subsection{Pitfalls}
\paragraph{Error sources on the human side} Based on our observed instability and expert feedback, we argue that the human cognitive errors, which either occur internally or are influenced by the system outcomes, are a crucial part of the overall system uncertainty:
\begin{itemize}[leftmargin=*]
\item[1)] \emph{Heuristic biases}.
a)~The \emph{anchoring bias} explains that earlier experience influences human decisions, including earlier system output and other context factors, such as background knowledge or expertise. In \ref{case:1} (see Section~\ref{sec:semi_structured_interviews}), the artist confirmed that his evaluation depended on meshes he had seen before.
b)~The \emph{availability bias} explains that judgments are based on the quickly accessed memories of relevant examples. The \ref{case:3} matches this bias as the artist decides based on his professional experience.
c)~\emph{Representativeness} shows that decisions made by substitution examples may be occasionally biased. The \ref{case:2} shows this behavior because the actual decision used mental shortcut and was made by judging another similar case.
\item[2)] \emph{Loss aversion and endowment effect}.
a)~Users may become more critical after observing several good results from an intelligent system. Users might stick to what they know and are familiar with and reject newly proposed and objectively good choices, which leads to more negative ratings later in the process. This may explain (\ref{case:3}) why artists stuck to mediocre choices in intermediate stages instead of moving to a broader (but more risky) range of variations.
b)~The software functionality (in our case, this is the software pre-configured camera angle for displaying meshes) as a task context may override information and influence the validity of the human judgment. This also explains the unexpected rating of the wheel model in \ref{case:3}.
c)~Human preferences change over time and may become inconsistent when interacting. A present rating choice also carries long-term influences, in contrast to being just local. In our case, this is explained by the anchoring bias. It was expected to be addressed in PBO, which uses comparative judgments, but as the three discussed cases show, artists still keep previous experience (either accumulated expertise or short-term outcomes) in mind, which changes their preferential choices.
\item[3)] \emph{Diminishing returns}. Judgments may lose precision and contain increasing noise after humans have seen increasingly or partially good results. Hence, preference exploitation may become less effective, and the HITL system can no longer benefit from human knowledge.
In our case, when artists had seen a certain number of increasingly better meshes, they were less sensitive (\ref{case:3}) to further improvements by the algorithm. In contrast, they even gave more critical scores for the occasional poor results.
\end{itemize}
\paragraph{Error sources on the machine side} The other part of overall system uncertainty comes from the underlying algorithm and is emphasized by user errors:
\begin{itemize}[leftmargin=*]
\item[1)] \emph{Stable preference assumption}. The system performance in a HITL system suffers from the model assumption, and the outcomes may be undesired due to an invalid optimization. We observed that human judgments produce strongly local, partially global, time- and context-dependent errors, even with permanent goal changes (\ref{case:1}). This violates the prerequisites of any optimization technique that assumes a unique and stable utility function, including PBO. More importantly, human judgment is a fragile function to optimize for, and the commonly used \emph{independent and identically distributed} (i.i.d.) assumption in these algorithms does not hold in reality for humans. In turn, we need to generally rethink basic assumptions and approaches in the design of HITL systems. We should be more explicit about under what circumstances they can be applied appropriately to detect and exploit changes in latent user preference distributions and systematic errors.
\item[2)] \emph{Complete preference assumption}. The underlying optimization still implicitly assumes a user always has a complete preference, meaning that users are deemed to be able to provide a rating to reflect their preference consistently. In the current design, users rate four models instead of requiring them to choose one of the best. This design can mitigate the completeness assumption violation, as selecting the best might not be possible if comparing objects are not entirely comparable and involves multiple optimizing objectives. However, as the optimization process continues, human raters may lose their preference for rating different models due to bounded rationality.
\end{itemize}

\subsection{Potential Countermeasures}
The heuristics are rather hard to detect by the machine since human ratings may not be entirely judged for consistency (otherwise, the machine could provide ratings on its own, entirely defying the idea of HITL systems). Nevertheless, we propose several design guidelines to at least mitigate different types of decision noise as discussed in~\autoref{sec:decision-error} thereby may be more parametrically guiding users in further optimization steps:

\begin{itemize}[leftmargin=*]
\item[1)] Reduce \emph{level noise}: Provide a timeline to include intermediate results saved by users and allow them to return to those earlier results for comparison. This could help the user to compare new results to known ones and support a more objective comparison across iterations. It could also reduce user frustration and fear of losing the achieved quality, thereby mitigating problems from loss aversion and violated system assumptions;
\item[2)] Reduce \emph{stable pattern noise}: Indicate the optimization intention to the user, such as current system steps regarding exploitation and exploration. This could better frame the current context, therefore, mitigate representativeness and availability bias by keeping users from judging based on earlier examples.
\item[3)] Reduce \emph{transient noise}: One approach could be to occasionally present results from earlier iterations and check for consistency, although this would also assume stable preferences and require more user iterations. Another approach could provide more assistive visualization by highlighting the mesh difference between iterations. This could further reduce user workload and help mitigate simple oversight and obvious mistakes when distracted by unchanged parts or overlooking changed parts.
\end{itemize}

\subsection{Limitations}
Our UI was consciously simplified to a minimum in order not to distract from judgment and in an attempt to avoid usage complexity and improve overall usability. In the field study, due to the limited number of users and to not further confuse users by silently changing system behavior, we did not run any forms of A/B testing.
Although the subjects could explore the quality of the entire mesh by features such as enabling the wireframe, this might still have been too restrictive and lacked information about the changes. Highlighting the crucial changes may be helpful, but it also lacks the ability to customize references to show the difference between different proposals in a sequential optimized workflow. In hindsight, we learned that it might be useful to let users specify which parts of the mesh led to a particular rating.
Next, we wanted to ensure the generalizability of our results and thus, selected five models with two wireframe representations. On the other hand, our results may still suffer from a selection bias in the models we used.
Lastly, conducting a simulated user study~\cite{antti2017abc} and designing further statistically verifying the decision biases might also be helpful to compare simulated human inputs with controlled noise and the actual decision behaviors.
\section{Summary and Future Work}

In this work, we discussed a HITL system where an optimization process in the background exploits sequential user choices to optimize the system's future outcomes iteratively. Our case study provides evidence of challenges to human-AI loops in practice, produced by mutual negative influences.
Based on interaction data in the field and lab and discussions with expert artists, we reflected on concrete influences that can break preference-optimized HITL systems, namely by 1)~human decision biases and noise, 2)~system capabilities to deal with them, and 3)~subsequent impact on future human inputs.

The findings provide some answers to our initial research questions: RQ1) The constraints of cognitive effects and the underlying algorithm, such as \emph{heuristic biases}, \emph{endowment effect}, \emph{diminishing return}, and \emph{violated system assumptions}, can be used to explain our empirical observations. Supporting polygon reduction tasks using the HITL strategy requires resolving these issues. RQ2) The observed constraints also apply in a similar HITL context, and we proposed descriptive UI design directions as promising countermeasures to prevent HITL system outcomes from being highly unstable and eventually non-satisfactory.

For future work, we expect to verify the proposed countermeasures in various scenarios and test for similar or different phenomena in other domains. We will also perform a systematic analysis of the basic building blocks that include other HITL systems and related cognitive factors as a foundation to inform guidelines for more error-tolerant HITL systems.

\section{Open Science}
\label{sec:open-source}

We encourage readers to reproduce and extend our results. Our system tools, technical specifications, anonymized datasets (without proprietary models and associated user ratings), and evaluation scripts are open-sourced that may be found in \url{https://changkun.de/s/infloop}.

\begin{acks}
The authors would like to thank our industrial partners Stefan Sigl, Marco Petrassi, and Marvin Juschus for sharing challenges in their daily 3D workflow; our colleague Kai Holländer for helpful discussions and feedback; to Feng Chen for executing the lab study, and to Prof. Eyke Hüllermeier and Karlson Pfannschmidt for useful discussions in preference learning. 3D mesh artifacts are provided courtesy of WAY Digital Solutions, Jeff H, Jose Olmedo, kenik, yarulesemel, and Stephan Thieme.
This work was supported by the \grantsponsor{1}{Bavarian IuK Program}{https://www.iuk-bayern.de/} (\grantnum{1}{IUK1805-0004//IUK577/002}). Work on this project is also partly funded by the Bavarian State Ministry of Science and the Arts and coordinated by the Bavarian Research Institute for Digital Transformation (bidt).
\end{acks}

\balance

\bibliographystyle{ACM-Reference-Format}
\bibliography{ref}

\end{document}